\documentclass[twocolumn,pre,showpacs]{revtex4}
\usepackage{graphicx}
\usepackage{amsmath}
\usepackage{amsfonts}
\usepackage{mathtools}

\begin{document}
\title{Information-theoretic vs. thermodynamic entropy production\\ in autonomous sensory networks}

\author{A. C. Barato, D. Hartich and U. Seifert}
\affiliation{ II. Institut f\"ur Theoretische Physik, Universit\"at Stuttgart, 70550 Stuttgart, Germany}

\parskip 1mm
\def\d{{\rm d}}
\def\Ps{{P_{\scriptscriptstyle \hspace{-0.3mm} s}}}
\def\MF{{\mbox{\tiny \rm \hspace{-0.3mm} MF}}}
\def\i{\text{\scriptsize $\cal{I}$}} 

\begin{abstract}
For sensory networks, we determine the rate with which they acquire information about the changing external conditions. Comparing
this rate with the thermodynamic entropy production that quantifies the cost of maintaining the network, we find that there is no universal bound
restricting the rate of obtaining information to be less than this thermodynamic cost. These results are obtained within a general bipartite model
consisting of a stochastically changing environment that affects the instantaneous transition rates within the system. Moreover, they  are illustrated 
with a simple four-states model motivated by cellular sensing. On the technical level, we obtain an upper bound on the rate of mutual information analytically and calculate 
this rate with a numerical method that estimates the entropy of a time-series generated with a simulation.

\end{abstract}

\maketitle
\section{Introduction}

Acquiring and processing information about the instantaneous state 
of the environment is a prerequisite for survival for any living 
system. Sensory and signal transducting networks have evolved to 
achieve this task under a variety of external conditions as, e.g., 
the work on bacteria like {\sl Escherichia coli} has demonstrated so 
beautifully \cite{ecoli,berg,bialekbook}. Maintaining any biochemical network, however, has a 
metabolic cost associated with its inherent non-equilibrium nature.
This fact prompts the question whether there is a relation between
the performance of the network and its free energy consumption. For
sensory adaptation in {\sl E. coli}, such a trade-off between adaptation
speed and accuracy, and the energy dissipation rate has recently been
found in a theoretical model and confirmed with experimental data \cite{nature}.

From a more information-theoretic perspective, the question alluded to
above can be formulated more generally as to whether there is any
quantitative relation between the rate with which such a network
acquires information about the ever changing environmental conditions
and the rate of entropy production associated with the functioning
of the network. The universal concept for quantifying one side of this
balance is mutual information, or, more precisely, the rate of
mutual information. In the context of sensing, this rate has been
introduced in \cite{tostevin09},  where it was explored for the special case of 
Gaussian input and output signals and various network motifs. Importantly, as pointed out in \cite{tostevin09}, because it takes temporal correlations into account
the rate of mutual information is different from the static mutual information, which has been the subject of several recent investigations in genetic regulatory networks \cite{tkacik11}.
On the other hand, calculating the rate of mutual information without the assumption of Gaussian statistics represents a main challenge.   

In this paper, we obtain an upper bound on the rate of mutual information for a general 
bipartite model consisting of an environment that switches stochastically
between an arbitrary number of states and a network for which the internal 
transition rates depend on the instantaneous state of the environment. 
This set-up has the advantage that the total system is in a non-equilibrium 
steady state generated by a Markovian dynamics which facilitates the analysis.
Still, since the internal process is non-Markovian, deriving the 
expression for this bound as done below becomes 
non-trivial. Moreover, in order to calculate the precise value of the rate of mutual information 
we apply a numerical method that estimates the entropy rate of a single time-series produced from
a numerical simulation \cite{holliday06,jacquet08,roldan12}.
For evaluating the second side of the balance introduced above, 
we need the rate of (thermodynamic) entropy production, as it has been 
derived for any network with given transition rates some time ago \cite{schnakenberg76} and 
revitalized in the context of stochastic thermodynamics as reviewed in \cite{seifert12}.

Significant progress in relating information-theoretic concepts to
thermodynamic ones has recently been achieved in the context of 
feedback-driven systems \cite{touchette00,cao09,sagawa10,horowitz11,esposito12,abreu12}. For these systems, the amount of mutual information between
system and controller enters the corresponding thermodynamic expressions
on the level of generalized fluctuation theorems and second-law like 
inequalities. In particular, the net-power output of such systems can not exceed 
the rate of mutual information acquired. The single cell, however, is an 
autonomous system for which a separation into an act of measurement and 
subsequent feedback control does not come naturally. Whether despite this fundamental difference between a 
feedback-driven system and an autonomous one, an
analogous constraint on a putative efficiency relating the rate of acquiring 
information with the thermodynamic cost of maintaining the sensory network exists
will be explored here.

The paper is organized as follows. In section \ref{sec2} we analyze the rate of mutual information in a simple toy model. The general framework, for which our results are valid, is introduced in section \ref{sec3}.
Section \ref{sec4} contains the comparison between the dissipation rate and the mutual information rate for a model of a cell computing an external ligand concentration. The rate of mutual information in equilibrium 
is discussed in section \ref{sec5}. We conclude in section \ref{sec6}. 


\begin{figure}
\includegraphics[width=72mm]{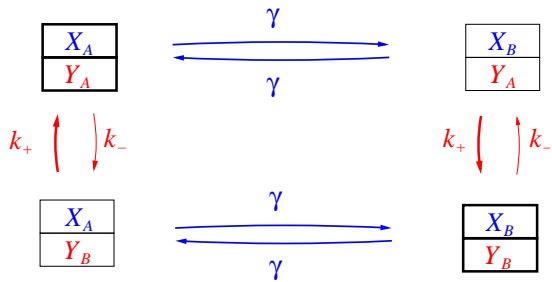}
\vspace{-2mm}
\caption{(Color online) Network and transition rates of a simple model of an internal process $Y$ coupled to an external one $X$.}
\label{fig1} 
\end{figure}

\section{Toy model}
\label{sec2}

In order to introduce this rate of mutual information between external and internal processes and compare it to the entropy production, we first consider the 
minimalistic model shown in Fig. \ref{fig1}. The environment is supposed to switch stochastically with rate $\gamma$ between two possible states $X_A$ and $X_B$. 
Within the system, there is an internal degree of freedom, which exists also 
in two different states that are correlated with the environment state. 
Specifically, if the external state is $X_A$ ($X_B$) and the internal state is in the 
``false'' state $Y_B$ ($Y_A$), the internal state changes to $Y_A$ ($Y_B$) with 
a rate $k_+$. We also allow for a (small) rate $k_-$ inducing ``false'' 
transitions. Later we will consider a more realistic model for bacterial chemotaxis,  where the external states correspond to the state of a receptor sitting on the membrane, while the internal states correspond to some internal protein that can be
transformed to an active form with transition rates that depend on the state of the receptor.    

As we will show below in equation (\ref{mutualcont}) using a general framework, the rate $\i$ at which the internal system acquires information about the time-dependent state of the environment
is bounded from above by
\begin{equation}
\i^{(u)} = f(k_+,k_-)+f(k_-,k_+),
\label{mutualtoy1}
\end{equation}  
where 
\begin{equation}
f(x,y)\equiv\frac{x(y+\gamma)}{2\gamma+x+y}\ln\frac{x(x+y+2\gamma)}{2xy+(x+y)\gamma}.
\end{equation}
For $k_-=0$ and in the limit $k_+\gg\gamma$, the rate of mutual information can be understood as follows. The uncertainty 
about the external time-series $X(t)$ still left after recording the internal time-series $Y(t)$
is basically the specific time lapse between a change in the 
external conditions and the change of the internal state, since for $k_-=0$ an internal jump takes place only after an external jump. 
Thus the system can localize the change in the external conditions within a time window of 
width $1/k_+$. Encoding the length of intervals between switching events on a 
scale $1/k_+$ requires a number of order $k_+/\gamma$, which carries 
$\ln (k_+/\gamma)$ units of information (measured throughout this paper using 
natural logarithms). Since these switching events take place with a rate 
$\gamma$, the rate of mutual information, for $k_-=0$ and $k_+\gg\gamma$, is $\i \simeq \gamma\ln\frac{k_+}{\gamma}$. Therefore, in this limit,
the upper bound (\ref{mutualtoy1}) gives the correct value of the rate of mutual information.  

\begin{figure}
\includegraphics[width=72mm]{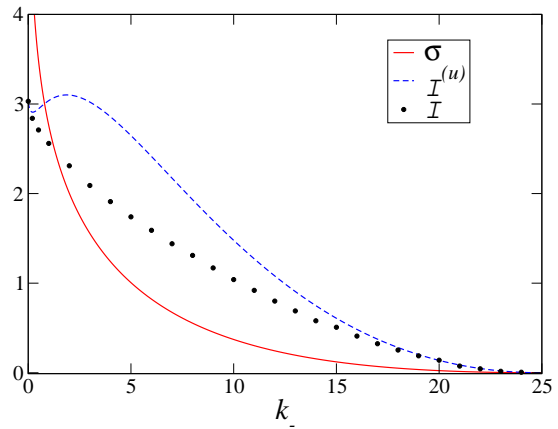}
\vspace{-2mm}\caption{(Color online) The rate of mutual information $\i$, estimated with the numerical method explained in the text, the upper bound $\i^{(u)}$,
and the thermodynamic entropy production $\sigma$ as a function of $k_-$ for $k_+= 25$ and $\gamma= 1$. Here and in the following figures the common unit of time is arbitrary and
the error in the numerics is less than the size of the symbols.
}
\label{fig2} 
\end{figure}

The dissipation rate $\sigma$ associated with this network follows from the standard expression for Markovian processes, recalled in equation (\ref{sdef}) below, which gives
\begin{equation}
\sigma= g(k_+,k_-)+g(k_-,k_+),
\end{equation}
where 
\begin{equation}
g(x,y)\equiv\frac{x(\gamma+y)}{2\gamma +x+y}\ln \frac{x}{y}.
\end{equation}
 The thermodynamic entropy production fulfills $\sigma\ge 0$ and is zero only if detailed balance is fulfilled ($k_-=k_+$). 
 
Whereas the rate of mutual information is finite at $k_-=0$, the entropy production diverges as $k_-\to 0$. Therefore, when the system is far from equilibrium ($k_-\ll k_+$) the dissipation
rate is larger than the mutual information rate. On the other hand, at equilibrium ($k_-=k_+$) both quantities are zero: there is no dissipation and the internal and external processes are uncorrelated. Note that in this
case also the internal process becomes Markovian. In Fig. \ref{fig2} we compare the rate of mutual information obtained from numerical simulations, as explained below, with the thermodynamic entropy production. 
There is a value of $k_-$, which depends on $k_+$ and $\gamma$, beyond which the mutual information rate becomes larger than
the thermodynamic entropy production. Thus, as our first main result, we have found that in the most simple conceivable model the rate of mutual information is not bounded
by the dissipation rate.
 

\section{General framework}
\label{sec3}

We now address the problem of calculating the upper bound on the rate of mutual information within a general bipartite model. It will be convenient to first treat time as discrete with a spacing $\tau$.
We denote the external states by $\alpha,\beta$ and their transition probabilities by $W^{\alpha\beta}\equiv w^{\alpha\beta}\tau$, where the transition is from $\alpha$ to $\beta$. The transition probability from the internal
state $i$ to the internal state $j$, given that the external state  is $\alpha$, is  $W_{ij}^\alpha \equiv w_{ij }^\alpha \tau$. Hence, the transition probabilities of the 
internal transitions depend on the instantaneous external state. States of 
the total system are thus determined by the pair $(\alpha,i)$, with the transition probabilities of this full Markov process given by
\begin{equation}	
W_{ij}^{\alpha\beta}\equiv\left\{
\begin{array}{l} 
 w^{\alpha\beta}\tau \quad \textrm{if $i=j$ and $\alpha\neq\beta$} \\
 w^{\alpha}_{ij}\tau \quad \textrm{if $i\neq j$ and $\alpha=\beta$}\\
 0 \quad \textrm{if $i\neq j$ and $\alpha\neq\beta$}\qquad. 
\end{array}\right.\,
\label{defrates}
\end{equation}
In the context of cellular sensing, the external transition rates $w^{\alpha\beta}$ are related to processes that happen outside the cell and that are not influenced by the biochemical reactions inside the cell. The internal biochemical 
reactions are related to the transition rates $w^{\alpha}_{ij}$, depending on the external state $\alpha$. A central consequence of the independence of $w^{\alpha\beta}$ on the internal state $i$ is that the external process
is also Markovian, while the internal process is in general non-Markovian.

A stochastic trajectory with $N$ jumps is denoted by $\{(X_t,Y_t)\}_{t=0}^{N\tau}$, where $\{X_{t}\}_{t=0}^{N\tau}$ ($\{Y_{t}\}_{t=0}^{N\tau}$) represents the external (internal) process. The information-theoretic entropy rate 
of the full process measures how much the entropy of the 
trajectory grows with $N$. In the limit $N\to\infty$, it is given by \cite{cover}
\begin{equation}
H_{X,Y}= -\frac{1}{\tau}\sum_{i,j,\alpha,\beta}P^\alpha_i W_{ij}^{\alpha\beta}\ln W_{ij}^{\alpha\beta},
\label{entropyxy}
\end{equation} 
where $P^\alpha_i$ denotes the stationary state probability distribution and, from equation (\ref{defrates}), the diagonal terms take the form 
\begin{equation}
W_{ii}^{\alpha\alpha}= 1-\sum_{\beta\neq\alpha}w^{\alpha\beta}\tau-\sum_{j\neq i}w^{\alpha}_{ij}\tau.
\end{equation}
Moreover, the external process $\{X_{t}\}_{t=0}^{N\tau}$ is also Markovian, which implies for its entropy rate
\begin{equation}
H_X= -\frac{1}{\tau}\sum_{\alpha,\beta}P^\alpha W^{\alpha\beta}\ln W^{\alpha\beta},
\label{entropyx}
\end{equation} 
where $P^{\alpha}=\sum_i P_i^\alpha$ and $W^{\alpha\alpha}=1-\sum_{\beta\neq\alpha}w^{\alpha\beta}\tau$.
Here, we are interested in calculating the rate of mutual information defined as \cite{cover}
\begin{equation}
I_{X,Y}= H_X+H_Y-H_{X,Y}.
\label{mutualdef}
\end{equation}

The still missing piece for evaluating this expression is the entropy rate of the internal process $H_Y$, which is not known
because $Y_t$ is in general non-Markovian. However, this quantity is bounded from above and from below by the relation \cite{cover}
\begin{equation}
H(Y_{n+1}|Y_{n},\ldots,Y_1,X_1)\le H_Y\le H(Y_{n+1}|Y_{n},\ldots,Y_1),
\label{bounds}
\end{equation}
which involves conditional entropies. Here, we are considering a finite sequence of $n$ jumps of the internal process starting with the stationary state probability distribution. These bounds become better as $n$ increases and for $n\to\infty$ 
both converge to the same value which is the entropy rate $H_Y$ \cite{cover}. The first upper bound is given by
\begin{equation}
H(Y_2|Y_1)= -\frac{1}{\tau}\sum_{Y_1} P(Y_1)\sum_{Y_2} P(Y_2|Y_1)\ln P(Y_2|Y_1).
\end{equation}
Using the transition rates (\ref{defrates}), for $Y_2\neq Y_1$, we have  
\begin{equation}
P(Y_2|Y_1)= \frac{\sum_{X_1}P(Y_2,Y_1,X_1)}{P(Y_1)}= \frac{\sum_\alpha P_i^{\alpha}w^{\alpha}_{ij}\tau}{P_i},
\end{equation}
where we substituted $X_1\to \alpha$, $Y_1\to i$, and $Y_2\to j$. With this relation it is easy to obtain
\begin{eqnarray}
H(Y_2|Y_1)=-\sum_{i,\alpha}P_i^\alpha\sum_{j\neq i}\left(w_{ij}^\alpha\ln\frac{\sum_\beta P_i^\beta w_{ij}^\beta \tau}{P_i}-w_{ij}^\alpha\right)\nonumber\\
+\textrm{O}(\tau).
\label{hy1}
\end{eqnarray}  
Moreover, from (\ref{entropyxy}) and (\ref{entropyx}), we get
\begin{equation}
H_{X}-H_{X,Y}= \sum_{i,\alpha}P_i^\alpha\sum_{j\neq i}\left(w_{ij}^\alpha\ln w_{ij}^\alpha\tau-w_{ij}^\alpha\right)+\textrm{O}(\tau).
\label{hxyhx}
\end{equation}  
Therefore, the first upper bound for the rate of mutual information, as obtained from (\ref{mutualdef}), (\ref{hy1}), and (\ref{hxyhx}), reads
\begin{equation}
I_{X,Y}^{(u,1)}=\sum_{i,\alpha}P_i^\alpha\sum_{j\neq i} w^\alpha_{ij}\ln \frac{w^\alpha_{ij}}{\overline{w_{ij}}}+\textrm{O}(\tau), 
\end{equation}
where 
\begin{equation}
\overline{w_{ij}}\equiv \sum_\beta P(\beta|i)w_{ij}^\beta
\end{equation}
is the mean internal transition rate and $P(\beta|i)= P_i^\beta/P_i$  is the probability of being in the external state $\beta$ given that the internal state is $i$. 
More generally, we find that up to $O(\tau \ln\tau)$ all upper bounds with finite $n$ are given by 
this expression, meaning that 
\begin{equation}
I_{X,Y}^{(u,n)}=\sum_{i,\alpha}P_i^\alpha\sum_{j\neq i} w^\alpha_{ij}\ln \frac{w^\alpha_{ij}}{\overline{w_{ij}}}+\textrm{O}(\tau\ln\tau).
\end{equation}
As for the lower bounds, we find that $I_{X,Y}^{(l,n)}= \textrm{O}(\tau\ln\tau)$ for all $n$. Therefore, we conclude that the rate of mutual information in the continuous time limit is bounded from above by
\begin{equation}
 \i^{(u)} = \sum_{i,\alpha}P_i^\alpha\sum_{j\neq i} w^\alpha_{ij}\ln \frac{w^\alpha_{ij}}{\overline{w_{ij}}}.
\label{mutualcont}
\end{equation}
We note that all the bare entropy rates diverge as $\ln \tau$ for $\tau\to 0$ and, therefore, cannot be defined in this limit \cite{gaspard04,lecomte07}. However, the rate of mutual information
is a well behaved quantity in this limit with no logarithmic divergences.

\begin{figure}
\includegraphics[width=72mm]{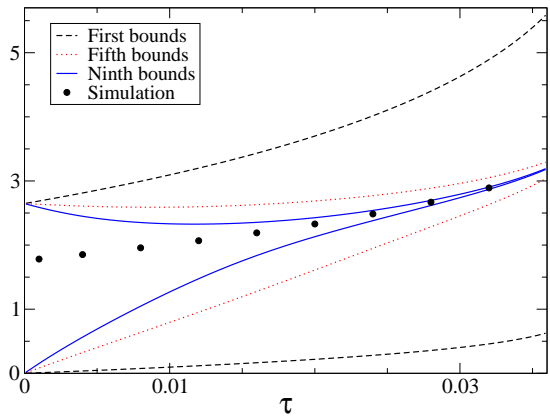}
\vspace{-2mm}
\caption{(Color online) The lower and upper bounds on the mutual information rate and the numerical simulation result as a function of $\tau$ for $k_-=5$, $k_+=25$, and $\gamma=1$. 
The simulations were done for a time series with $N=10^{10}$.}
\label{fig3} 
\end{figure}

Let us now explain the numerical method we use to estimate the entropy rate $H_Y$ of the non-Markovian time series $\{Y_{t}\}_{t=0}^{N\tau}$. The random matrix $\mathcal{T}(Y_t,Y_{t-1})$,
is defined as $\mathcal{T}(Y_t,Y_{t-1})_{X_t,X_{t-1}}= P[Y_t,X_t|Y_{t-1},X_{t-1}]$, where $P[Y_t,X_t|Y_{t-1},X_{t-1}]$ is a conditional probability.
If the external states take the values $X_t= 1,\ldots,\Omega$, then $\mathcal{T}$ is a $\Omega\times\Omega$ matrix,
where the internal variables $(Y_t,Y_{t-1})$ make the matrix random. It is simple to show that \cite{jacquet08,holliday06}
\begin{equation}
P[\{Y_{t}\}_{t=0}^{N\tau}]=  \mathcal{V} \mathcal{T}(Y_N,Y_{N-1})\ldots \mathcal{T}(Y_1,Y_{0}) \mathcal{P}_0
\end{equation}
where $P[\{Y_{t}\}_{t=0}^{N\tau}]$ is the probability of the time series, $\mathcal{V}$ is a vector with all components equal to one and $\mathcal{P}_0$ is a vector with components $P(Y_0,X_0)$, for $X_0= 1,\ldots,\Omega$.
From this relation, it is then possible to show that given a typical long time series  $\{y_{t}\}_{t=0}^{N\tau}$, one can estimate the entropy rate $H_Y$ by the formula \cite{jacquet08,holliday06,roldan12}
\begin{equation}
H_Y\approx \frac{1}{N\tau}\ln \lVert \prod_{t=1}^{N}\mathcal{T}(y_t,y_{t-1})\rVert,
\end{equation}
where $ \lVert\cdot \rVert$ is any matrix norm. Therefore, following Ref. \cite{roldan12}, we can calculate $H_Y$ numerically by generating an internal time-series with a simulation and calculating
the above product by normalizing it after a certain number of steps (keeping track of the normalization factors), as explained in \cite{crisanti93}. In Fig. \ref{fig3}, we plot upper and 
lower bounds on the mutual information rate, obtained from (\ref{bounds}), and the numerical
simulation result. For small $\tau$, the numerically obtained rate of mutual information shows linear behavior,
which we extrapolate to $\tau\to 0$ to estimate the rate of mutual information in the continuous time limit.

The other central physical observable we consider here is the thermodynamic entropy production \cite{schnakenberg76,seifert12,lebowitz99}, which for the present class of Markov processes reads
\begin{equation}
\sigma= \sum_{\alpha,\beta}P^\alpha w^{\alpha\beta}\ln \frac{w^{\alpha\beta}}{w^{\beta\alpha}}+\sum_{\alpha}\sum_{i,j}P_i^\alpha w^\alpha_{ij}\ln\frac{w^\alpha_{ij}}{w^\alpha_{ji}},
\label{sdef}
\end{equation}      
where the first term on the right hand side is due to the external transitions and the second is the contribution from internal transitions. If the same pair of internal states can be connected by different types of transitions, as
in the model discussed next, the second term requires an additional summation over the different ``channels'' $\nu$ with corresponding rates $w_{ij}^{\alpha(\nu)}$ \cite{schnakenberg76}. 

\section{Thermodynamically consistent model}
\label{sec4}

We finally use this framework to analyze a thermodynamically consistent
minimal four-states model for a cell estimating a ligand concentration $c$ \cite{berg,bialekbook,bialek05,endres09,mora10,mehta12}). 
The external states are related to a receptor that can be either bound by a ligand or empty. Moreover, the receptor can be
in an ``on'' state or ``off'' state. We assume a high binding affinity of the on state, i. e., 
whenever the receptor is occupied by an external ligand it is in the on state. Likewise, any unbound receptor is in the off state.
Under these simplifying conditions the state of the receptor (on/off) corresponds to what was called above the external process $X_t$. 
The transition from on to off
happens at a rate $k_{\textrm{off}}$, whereas the binding rate $k_{\textrm{on}}$ is 
proportional to the ligand concentration $c$. The internal state 
is associated with a downstream protein that can be in an inactive ($Y$) 
or, due to phosphorylation, in an active state $Y^*$. For simplicity, we assume a two states internal system corresponding to one downstream protein. In reality,
the number of downstream proteins is much larger than one \cite{mehta12}, however, this simplification is not harmful for the qualitative comparison between the mutual 
information rate and thermodynamic entropy production.  

\begin{figure}
\includegraphics[width=72mm]{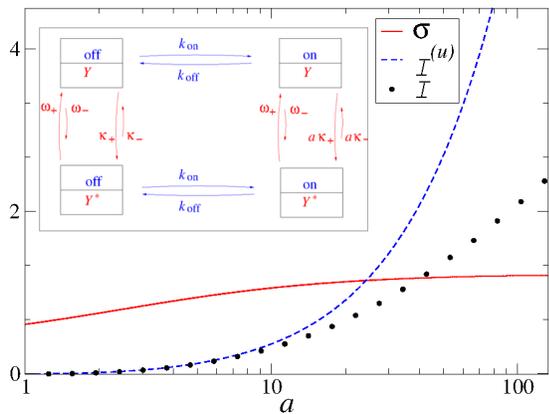}
\vspace{-2mm}
\caption{(Color online) The entropy production rate $\sigma$, the upper bound $\i^{(u)}$ and the numerically determined rate of mutual information $\i$ as a function of $a$ 
for the thermodynamically consistent model shown in the inset. The other parameters are set to $k_{\textrm{on}}=k_{\textrm{off}}=1$, $\kappa_-=1/10$, $\kappa_+=4/10$, $\omega_-=1/10$, and $\omega_+=6/10$.}
\label{fig4} 
\end{figure}

The crucial coupling of the internal process to the instantaneous state of the environment, here encoded by 
the state of the receptor, arises from the fact that the receptor in the on state
speeds up phosphorylation, which happens at a rate $\kappa_+$, by a factor $a\ge1$ compared to the 
action of an empty receptor. Dephosphorylation occurs at rate $\omega_+$, which leads to the following internal ATP consumption cycle 
\begin{equation}
 Y+ATP\xrightleftharpoons[(a)\kappa_-]{(a)\kappa_+} Y^*+ADP \xrightleftharpoons[\omega_-]{\omega_+} Y+ADP+P_i,
\label{eqreaction} 
\end{equation}
where the factor $a$ arises only if the receptor is in the on state. The full network of transitions is shown in Fig. \ref{fig4}.
Thermodynamic consistency requires, first, that we also allow for the reverse transitions of 
phosphorylation and dephosphorylation with non-zero rates $(a)\kappa_-$ and
$\omega_-$, respectively. Second, it imposes a relation between the free energy associated with the ATP hydrolysis $\Delta \mu$ and the kinetic rates, which reads 
\begin{equation}
\Delta \mu= k_BT\ln \frac{\kappa_+\omega_+}{\kappa_-\omega_-},
\end{equation}
 where $k_B$ is Boltzmann's constant and $T$ the temperature \cite{schnakenberg76, seifert12}.

In Fig. \ref{fig4}, we compare the rate of mutual information, obtained by the numerical simulation with the rate
of free energy consumption $\sigma$ as a function of the enzymatic enhancement $a$. For $a=1$, the internal
state is decorrelated from the external one, leading to $\i =0$. However, if $\Delta \mu\not
=0$, the internal network still consumes free energy. With increasing
$a$ the rate of mutual information increases, eventually becoming larger
than the rate of free energy consumption. Thus, this thermodynamically
fully consistent model confirms what we have found previously in the
simple model: in such autonomous networks the rate of mutual information
is not bounded by the free energy consumption required to maintain the
network.

\section{Mutual information in equilibrium networks}
\label{sec5}

While a cell undoubtedly is a nonequilibrium system, it is instructive to explore whether mutual information can be non-zero even under equilibrium conditions, i.e.,
for transition rates such that the steady state fulfills detailed balance. Two cases must be distinguished.

First, we stay within our framework of Markov processes of the form (\ref{defrates}), where the external transition rates $w^{\alpha\beta}$ are independent of the internal state $i$. A non-zero rate of
mutual information in equilibrium can then occur if the external process only affects the time-scale of the internal transition rates, i.e., if for all $i,j$ the ratio $w^\alpha_{ij}/w^\alpha_{ji}$ is independent of $\alpha$
and if $w^\alpha_{ij}= w^\beta_{ij}$ does not hold for  all $\alpha,\beta$. As an example, consider the model of Fig. \ref{fig1} with $k_+=k_-$ and the left vertical 
transition rates for $X=X_A$ multiplied by a factor $r$. This is clearly
an equilibrium model obeying detailed balance and for $r\neq 1$ the rate of mutual information is not zero. 

The above example is a rather contrived case where the external states distinguish between internal processes that differ only by the time-scale of the transitions. The question 
we raise in this paper, whether the dissipation rate bounds the rate of mutual information, is non-trivial for Markov processes of the form (\ref{defrates}), for which the external process truly affects the internal process and for which if
detailed balance is fulfilled in a point of the phase diagram then all internal processes are exactly the same, i.e., $w_{ij}^{\alpha}\equiv w_{ij}$. In this case, at equilibrium the internal process becomes also 
Markovian and decoupled from the external process, implying a zero rate of mutual information. Hence, it is not clear a priori whether the rate of mutual information is bounded by the dissipation rate under
nonequilibrium conditions.

Second, a different situation arises if one gives up the condition that the external process $X_t$ is unaffected by the internal process $Y_t$. By dropping this distinction,
a general pair of stochastic variables that together fully specify an underlying Markov process fulfilling detailed balance might have a non-zero rate of mutual information.  
An important example in the context of chemotaxis is the Monod-Wyman-Changeux (MWC) model \cite{bialekbook}. This model describes the allosteric interaction between the kinase activity of the 
receptor (on/off) and its affinity for ligand binding (bound/unbound). This aspect can be made explicit also with the toy model of Fig. \ref{fig1}. If   
we modify the transition rates from $X_B$ to $X_A$ when $Y=Y_A$ by a factor $k_+/k_-$ and 
when $Y=Y_B$ by an factor $k_-/k_+$ (see \cite{tkacik11} for a similar model), we get an equilibrium MWC like model which should have a non zero rate of mutual information between $X_t$ and $Y_t$.
For this variant, it is not possible to distinguish 
an external process influencing the internal process but not being influenced by it. More precisely, the transition rates are no longer of the form (\ref{defrates}) with $w^{\alpha\beta}$ independent of the internal state $i$.
While it would be interesting to calculate the rate of mutual information for the MWC model, our framework based on the rates of the form (\ref{defrates}) is not yet appropriate to do so. In general, our framework is
suited to study the rate of mutual information involving an external process that influences the chemical reactions inside the cell but is not affected by them.

\section{Conclusion}
\label{sec6}

In this final section we first come back to a topic raised in the introduction, namely how our main result, i.e.,
no bound between the rate of mutual information and dissipation, relates to the apparently quite different 
results for feedback-driven systems. In the latter,
information acquired through a measurement is used to extract net work from
a thermal bath. The amount of net work is limited by the information. Here,
for the sensory network, we have investigated a complementary issue,
namely whether the amount of information is limited by the chemical work, or free 
energy consumption, required to maintain the network. How this information is now used 
in a second step for an action that possibly performs work on some other 
element is an interesting but quite different question. As an important
aside, we note that for an autonomous network the whole issue of writing 
the information into a memory whose erasure will require free energy \cite{maruyama09,mandal12} is 
irrelevant as the erasure process is trivially included in the reaction scheme.

Clearly, these concluding remarks touch on deep issues concerning a 
future theory comparing comprehensively autonomous with feedback-driven
systems which is a distinction that may become blurred in the micro
world. On a more specific level, our study should now be refined by
incorporating further elements of a more elaborate model of a
sensory network, i.e., including several receptors, several proteins, 
allowing for adaptation and alike. The quantitative balance between 
information rate and entropy production will depend on such details, and 
thus, e.g., optimization of the ``efficiency'' will become an interesting 
issue. Searching for a hard universal thermodynamic 
bound for it, however, should be futile, as our simple model shows. Furthermore,
it would be worthwhile
to investigate the relation between the rate of mutual information, the dissipation rate and
the adaptation error in the context of the energy-speed-accuracy relation found in \cite{nature}.
 Finally, an exact calculation
of the rate of mutual information which would replace the upper bound we have derived here remains an open mathematical challenge.    

\begin{acknowledgments}
Support by the ESF though the network EPSD  is gratefully acknowledged. 
\end{acknowledgments}


\end{document}